\documentclass{bioinfo}
\copyrightyear{2013}
\pubyear{2013}
\usepackage{algorithm}
\usepackage{algorithmic}
\usepackage{url}
\usepackage{subfigure}
\newtheorem{definition}{Definition}

\newtheorem{theorem}{Theorem}[section]

\newcommand{\nop}[1]{}

\begin{document}
\firstpage{1}

\title[MSPKmerCounter]{MSPKmerCounter: A Fast and Memory Efficient Approach for K-mer Counting}
\author[Li]{Yang Li and Xifeng Yan}
\address{Department of Computer Science, University of California at Santa Barbara}

\history{Received on XXXXX; revised on XXXXX; accepted on XXXXX}

\editor{Associate Editor: XXXXXXX}

\maketitle

\begin{abstract}

\section{Motivation:}
A major challenge in next-generation genome sequencing (NGS) is to assemble massive overlapping short reads that are randomly sampled from DNA fragments. To complete assembling, one needs to finish a fundamental task in many leading assembly algorithms: counting the number of occurrences of k-mers (length-k substrings in sequences). The counting results are critical for many components in assembly (e.g. variants detection and read error correction). For large genomes, the k-mer counting task can easily consume a huge amount of memory, making it impossible for large-scale parallel assembly on commodity servers.

\section{Results:}
In this paper, we develop MSPKmerCounter, a disk-based approach, to efficiently perform k-mer counting for large genomes using a small amount of memory. Our approach is based on a novel technique called Minimum Substring Partitioning (MSP). MSP breaks short reads into multiple disjoint partitions such that each partition can be loaded into memory and processed individually. By leveraging the overlaps among the k-mers derived from the same short read, MSP can achieve astonishing compression ratio so that the I/O cost can be significantly reduced. For the task of k-mer counting, MSPKmerCounter offers a very fast and memory-efficient solution. Experiment results on large real-life short reads data sets demonstrate that MSPKmerCounter can achieve better overall performance than state-of-the-art k-mer counting approaches.

\section{Availability:}
MSPKmerCounter is available at http://www.cs.ucsb.edu/ $\sim$yangli/MSPKmerCounter

\section{Contact:} \href{yangli@cs.ucsb.edu}{yangli@cs.ucsb.edu}
\end{abstract}

\section{Introduction}

High-quality genome sequencing plays an important role in genome research.\nop{Recently, massively parallel DNA sequencing technologies (\citealp{Mardis08}), are improving continuously, making it possible for researchers to obtain enough sequence reads to try the assembly of higher eukaryotes at ever lowering costs.} A central problem in genome sequencing is assembling massive short reads generated by the next-generation sequencing technologies (\citealp{Mardis08}). These reads are usually randomly extracted from samples of DNA segments. Typically a modern technology can produce billions of short reads whose length varies from a few tens of bases to several hundreds.  For example, massively parallel sequencing platforms, such as Illumina (www.illumina.com), SOLiD (www.appliedbiosystems.com), and 454 Life Sciences (Roche) GS FLX (www.roche.com), can produce reads from 25 to 500 bases in length.  The short read length is expected to further increase in the following years.

Despite the progress in sequencing techniques and assembly methods in recent years, de novo assembly remains a computationally challenging task. The existing de novo assembly algorithms can be classified into two main categories based on their internal assembly model: (1) The overlap-layout-consensus model, used by Celera (\citealp{Myers00}), ARACHNE (\citealp{Batzoglou02}), Atlas (\citealp{Havlak04}), Phusion (\citealp{Mullikin03}) and Forge (\citealp{Platt10}); (2) The de Bruijn graph model, used by Euler (\citealp{Pevzner01}), Velvet (\citealp{Zerbino08}), ABySS (\citealp{Simpson09}), AllPaths (\citealp{Butler08}) and  SOAPdenovo (\citealp{Li10a}). The overlap-layout-consensus model builds an overlap graph between reads. Since each read can overlap with many other reads, it is more useful for sequencing data sets with a small number of long reads.  The de Bruijn graph approach breaks short reads to k-mers (substring of length k) and then connects k-mers according to their overlap relations in the reads. The de Bruijn graph approach is usually able to assemble larger quantities (e.g., billions) of short reads with greater coverage.  Systematic comparison of these algorithms is given by \cite{Earl11} and \cite{Salzberg12}.

Although the de Bruijn graph approach comes up with a good framework to reduce the computation time for assembly, the graph size can be extremely large, for example, containing billions of nodes (k-mers) for genomes of higher eukaryotes like mammals. Therefore, large memory consumption is a pressing practical problem for the de Bruijn graph based approach (\citealp{Miller10}). For the short read sequences generated from mammalian-sized genomes, software like Euler, Velvet, AllPaths and SOAPdenovo will not be able to finish assembling successfully within a reasonable amount of memory. Due to this drawback, it significantly limits the opportunity to run de novo assembly on numerous commodity machines in parallel for large-scale sequence analysis. This problem has also blocked other application of de Bruijn graphs, e.g., variants discovery in \cite{Iqbal12}.

To deal with the memory issue, an error correction step is often taken to eliminate erroneous k-mers before constructing the de Bruijn graphs.  In most NGS data sets, a large fraction of k-mers arise from sequencing errors.  These k-mers have very low frequencies.  In the giant panda genome sequencing experiment (\citealp{Li10b}), the error correction process could eliminate 68\% of the observed 27-mers, reducing the total number of distinct 27-mers from 8.62 billion to 2.69 billion. Though error correction is usually helpful, obtaining the k-mer frequencies itself is a computationally demanding task for large genome data sets. One ``naive" solution is using a hash table, where keys are the k-mers and values are the corresponding k-mer frequencies.  Unfortunately, this approach will easily blow up main memory. For example, in the Asian genome short read data set (\citealp{Li10a}), if $k = 25$, there are about 14.6 billion distinct k-mers. Assuming a load factor of $2/3$ for the hash table and encoding each nucleotide with 2 bits, the k-mers table would require nearly 160 GB memory. Furthermore, this problem will become severe when the length of short reads produced by the next-generation sequencing techniques further increases.

A recently developed program called Jellyfish (\citealp{Marcais11}) is designed to count k-mers in a memory efficient way. It adopts a ``quotienting" technique to reduce the memory consumption of k-mers stored in a hash table. Implemented with a multi-threaded, lock-free hash table, it is able to count k-mers up to 31 nucleotides in length using a much smaller amount of memory than the previous ``naive" method. When there is no enough memory to carry out the entire computation, Jellyfish will write intermediate counting results to disk and later merge them. Since the same k-mer may appear in several different intermediate results, the merge operation is not just a simple concatenation process; it can be quite slow. Another state-of-the-art k-mer counting algorithm, BFCounter (\citealp{Melsted11}) is based on bloom filter, a probabilistic data structure that can also reduce memory footprint. \nop{It uses less memory than Jellyfish when the main memory is large enough for Jellyfish to carry out the entire computation without writing and merging intermediary results.} However, BFCounter is 3 times slower than Jellyfish when Jellyfish is able to finish the task in memory (\citealp{Melsted11}). And moreover, it might miss some counts.

In this paper, we develop MSPKmerCounter, a disk-based approach, to efficiently perform k-mer counting for large genomes using a small amount of memory. Our approach is based on a recently proposed technique called Minimum Substring Partitioning (MSP) (\citealp{Li13}). MSP breaks short reads to ``super k-mers" (substring of length greater than or equal to k) such that each ``super k-mer" contains k-mers sharing the same minimum p-substring ($p\leq k$). The effect is equivalent to compressing consecutive k-mers using the original sequences. It is shown that this compression approach does not introduce significant computing overhead, but could lead to partitions 10-15 times smaller than the direct approach using a hash function (\citealp{Li13}), thus greatly reducing I/O cost.

For the task of k-mer counting, MSPKmerCounter offers a very fast and memory-efficient solution. Experiment results on large real-life short reads data sets demonstrate that MSPKmerCounter can achieve better overall performance than state-of-the-art k-mer counting approaches like Jellyfish and BFcounter.

\section{Background}
\begin{definition}[Short Read, K-Mer]
A short read is a string over alphabet $\Sigma = \{A, C, G, T\}$ (in DNA assembly).   A $k$-mer is a string over $\Sigma$ whose length is $k$. Given a short read $s$, $s[i,j]$ denotes the substring of $s$ from the $i_{th}$ element to the $j_{th}$ element (both inclusive).  $s$ can be broken into $n-k+1$ $k$-mers, written as $s[1, k]$, $s[2, k+1], \ldots$, $s[n-k+1, n]$. Two k-mers in $s$, $s[i, k+i-1]$, $s[i+1, k+i]$ are called adjacent in $s$.
\end{definition}

We can view k-mers generated in a way that a window with width k slides through a short read $s$. The adjacency relationship exists between each pair of k-mers for which the last k-1 bases of the first k-mer are exactly the same as the first k-1 bases of the last k-mer. .

\begin{definition}[Reverse Complement]
DNA sequences can be read in two directions: forwards and backwards with each nucleotide changed to its Watson-Crick complement ($A \leftrightarrow T$  and $C \leftrightarrow G$). For each DNA sequence, its corresponding read in the other direction is called reverse complement and they are considered equivalent in bioinformatics.
\end{definition}

In most sequencing technologies, the fragments (short reads) are randomly extracted from the DNA sequence in either direction. Therefore, if two k-mers, $K_1$ and $K_2$, are adjacent from $K_1$ to $K_2$ in the short reads data set, it implies that the reverse complement k-mer of $K_2$, say $K_2$' and the reverse complement k-mer of $K_1$, say $K_1$', are adjacent from $K_2$' to $K_1$'. So in an assembly processing, each short read should be read twice, once in forward direction and then in the reverse complement direction. However, in real implementation, it is possible to avoid reading sequences twice by inferring the subgraph introduced by reverse complements later from the forward direction subgraph.

\section{Minimum Substring Partitioning}
Our approach to do fast and memory efficient k-mer counting is based on a disk-based partition approach called Minimum Substring Partitioning (MSP) (\citealp{Li13}). MSP is able to partition k-mers into multiple disjoint partitions, as well as retaining adjacent k-mers in the same partition. This nice property introduces two advantages: first, instead of being outputted as several individual k-mers, consecutive k-mers can be compressed to ``super k-mers" (substring of length greater than or equal to k), which will greatly reduce the I/O cost of partitioning; second, with adjacent k-mers in the same partition, it is possible to do local assembly for each partition in parallel and later merge them to generate the global assembly.

\begin{definition} [Substring]
A substring of a string $s=s_1 s_2 \ldots s_n$ is a string $t=s_{i+1} s_{i+2} \ldots s_{i+m}$, where $0 \leq i $  and $i+m\leq n$.
\end{definition}

\begin{definition} [Minimum Substring]
Given a string $s$, a length-p substring $t$ of $s$ is called the minimum p-substring of $s$, if $\forall s'$, $s'$ is a length-p substring of $s$, s.t., $t\leq s'$ ($\leq $ defined by lexicographical order). The minimum p-substring of $s$ is written as $min_{p}(s)$.
\end{definition}

\begin{definition}[Minimum Substring Partitioning]
\label{def:msp}
Given a string $s=s_1 s_2 \ldots s_n$, $p, k \in N$, $p\leq k \leq n$, minimum substring partitioning breaks $s$ to substrings with maximum length $\{s[i, j]| i+k-1\leq j, 1\leq i, j\leq n\}$, s.t., all k-mers in $s[i,j]$ share the same minimum p-substring, and it is not true for $s[i, j+1]$ and $s[i-1, j]$.  $s[i,j]$ is also called ``super k-mer".
\end{definition}

\begin{figure}[!tpbh]
\centerline{\includegraphics[width=0.37\textwidth]{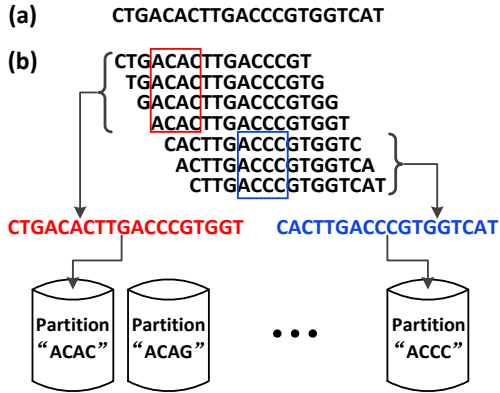}}
\caption{A minimum substring partitioning example: (a)short read (b)K-mers and MSP process}\label{fig:msp}
\end{figure}

Minimum Substring Partitioning comes from the intuition that two adjacent k-mers are very likely to share the same minimum $p$-substring if $p << k$, since there is a length-(k-1) overlap between them. Figure \ref{fig:msp} shows a Minimum Substring Partitioning example. In this example, the first $4$ k-mers have the same minimum $4$-substring, $ACAC$, as highlighted in red box; and the last $3$ k-mers share the same minimum $4$-substring, $ACCC$, as highlighted in blue box. In this case, instead of generating all these $7$ k-mers separately, we can just compress them using the original short read. Namely, we compress the first $4$ k-mers to $CTGACACTTGACCCGTGGT$, and output it to the partition corresponding to the minimum $4$-substring $ACAC$. Similarly, the last $3$ k-mers are compressed to $CACTTGACCCGTGGTCAT$ and outputted to the partition corresponding to the minimum $4$-substring $ACCC$. Generally speaking, given a short read $s=s_1 s_2 \ldots s_n$, if the adjacent $j$ k-mers from $s[i, i+k-1]$ to $s[i+j-1, i+j+k-2]$ share the same minimum $p$-substring $t$, then we can just output substring $s_i s_{i+1} \ldots s_{i+j+k-2}$ to the partition corresponding to the minimum $p$-substring $t$ without breaking it to $j$ individual k-mers. If $j$ is large, this compression strategy will dramatically reduce the I/O cost.

The results of the Minimum Substring Partitioning are determined by the parameters $k$ and $p$. Smaller $p$ will increase the probability that consecutive k-mers share the same minimum p-substring and thus reduce the I/O cost. However, it will also introduce a problem where the distribution of partition sizes become skewed and the largest partition may not fit in the main memory.  In the extreme case of $p = 1$, the size of the largest partition is almost as same as the size of the short reads data set and other partitions are almost empty (assuming the four nucleotides A, C, G, T are distributed randomly in the data set).  In that case, we lose the point of partitioning. On the other hand, larger $p$ will make the distribution of partition sizes evener at the cost of decreasing the probability that consecutive k-mers share the same minimum p-substring and thus increasing the I/O cost.  In the extreme case of $p \rightarrow k$, almost no adjacent k-mers will share the same minimum p-substring and thus no compression can be gained.  Therefore one needs to make a tradeoff (by varying $p$) between the largest partition's size and the I/O overhead. Fortunately, there is a quite wide range of values that $p$ can choose without affecting the performance of MSP (\citealp{Li13}).

\begin{definition}[Wrapped Partitions]
Given a string set $\{s_i\}$,  a hash function $H$, the user-specified number of partitions $N$, for any k-mer $s_{i,j}$, minimum substring partition wrapping assigns $s_{i,j}$ to the ($H(min_{p}(s_{i,j})) \mod N$)-th partition.
\end{definition}

Since each $p$-substring corresponds to one partition,  the total number of partitions in MSP is equal to $4^p$.  When $p$ increases, the number of partitions will increase exponentially and many partitions may become empty.  To address this problem, one can introduce a hash function to wrap the number of partitions to any user-specified partition number.  Then the k-mers are likely to be evenly distributed across partitions.

\begin{definition}[Minimum Substring with Reverse Complement]
\label{def:reverse}
Given a string $s$, a length-p substring $t$ of $s$ is called the minimum p-substring of $s$, if $\forall s'$, $s'$ is a length-p substring of $s$ or $s$' reverse complement, s.t., $t\leq s'$ ($\leq $ defined by lexicographical order).
\end{definition}

Definition \ref{def:reverse} redefines minimum substring by considering the reverse complement. With this new definition, we can make sure each k-mer and its reverse complement k-mer are assigned to the same partition. This property can help us save much time and memory in the later processing (e.g. storing only the lexicographical smaller one of a k-mer and its reverse complement k-mer in hash table and avoiding reading each short read twice to explicitly process reverse complement) since a k-mer and its reverse complement are considered equivalent in bioinformatics and the information introduced by reverse complement can be inferred from the forward direction short reads. For simplicity reason, in the following discussions, if not mentioned explicitly, we will ignore the reverse complement issue. However, in our implementation and experiments, we do consider its impact.

\section{Methods}
In this section, we describe the detailed method to do k-mer counting with the adoption of the minimum substring partitioning technique introduced in the last section.

The first step is to partition short reads. In this step, we will cut each short read of length $n$ into $(n-k+1)$ k-mers and then dispatch these k-mers into different partitions. The Minimum Substring Partitioning technique introduced in Section 3 is used as our partitioning method. As mentioned before, with this partitioning method, we can compress consecutive k-mers dispatched to the same partition into one ``super k-mer" to minimize the I/O cost.

There are several ways (e.g. straightforward, min-heap) to implement the minimum substring partitioning. Here we adopt the one introduced in \cite{Li13}, since it is proved to have the best performance in practice. The details of this implementation is described in Algorithm \ref{algo:partition}.

\begin{algorithm}
\caption{Minimum Substring Partitioning}
\label{algo:partition}
\begin{algorithmic}
\STATE Input: String $s=s_1 s_2 \ldots s_n$, integer $k, p$.
\STATE min\_s = the minimum p-substring of $s[1, k]$
\STATE min\_pos = the start position of min\_s in $s$

\FORALL {$i$ from $2$ to $n-k+1$}
    \IF {$i$ $>$ min\_pos}
        \STATE min\_s = the minimum p-substring of $s[i, i+k-1]$
        \STATE update min\_pos accordingly
    \ELSE
        \IF {the last p-substring of $s[i, i+k-1]$ $<$ min\_s}
            \STATE min\_s = the last p-substring of $s[i, i+k-1]$
            \STATE update min\_pos accordingly
        \ENDIF
    \ENDIF
\ENDFOR
\end{algorithmic}
\end{algorithm}

As mentioned before, we can view k-mers generated in a way that a window with width k slides through a short read. In Algorithm \ref{algo:partition}, initially when the window starts at position $1$, we scan the window to find the minimum p-substring, say min\_s, and the start position of min\_s, say min\_pos. Then we slide the window forward, one symbol each time, till the right bound of the window reaches the end of the short read.  After each sliding, we test whether the min\_pos is still within the range of the window. If not, we have to re-scan the window to get new min\_s and min\_pos. Otherwise, we test whether the last p-substring of the current window is smaller than current min\_s. If yes, we set this last p-substring as new min\_s and update min\_pos accordingly. If not, we just keep the old min\_s to calculate the partition location. As described in last section, the neighboring k-mers will likely contain the same minimum p-substring.  Therefore, the re-scan of the whole window will not occur very often.   The worst case time complexity is $O(nk)$ $p$-substring comparisons.  However, this algorithm is more efficient in practice (close to $O(n+lk)$, see detailed proof in \cite{Li13}) when $s$ is broken to only a few number ($l$) of ``super k-mers".  This is very true in minimum substring partitioning of real short reads. Table \ref{Tab:00} shows that the average number of breakdowns is small for several real short reads data sets.

\begin{table}[!th]
\processtable{Average number of breakdowns for real short reads data sets\label{Tab:00}}
{\begin{tabular}{lcccc}\toprule
Data Set &$n$ &$k$ &$p$ &Average Breakdown ($l$) \\\midrule
Budgerigar &150 &59 &10 &5.22 \\
Red tailed boa constrictor &121 &59 &10 &3.89 \\
Lake Malawi cichlid &101 &59 &10 &2.77 \\
Soybean &75 &59 &10 &1.69 \\\botrule
\end{tabular}}{}
\vspace{-3mm}
\end{table}

Note that in Algorithm \ref{algo:partition}, every time when we capture a minimum substring change at position $j$ of $s$ or we reach the end of $s$, we output a ``super k-mer" of $s$ that contains the previous minimum substring into the partition corresponding to that minimum substring. This part of code is not presented in Algorithm \ref{algo:partition}.

After obtaining the partitions, we can use a simple hash table whose keys are k-mers and values are k-mer counts to count the k-mer frequencies. For each partition, break the ``super k-mers" into k-mers and insert these k-mers into a hash table. Since adjacent k-mers are only different by the first and last symbol, direct bit shift operations (A, C, G, T can be encoded using 2 bits) can be applied here to improve the efficiency. Whenever we see a new k-mer, we first look up the hash table to see if it is already in the hash table: if yes, we increase the frequency count by 1; otherwise, we put this k-mer along with an initial frequency value 1 into the hash table. After processing one partition, write the entries in hash table to a disk file\footnote{Actually we will sort the k-mers in hash table before writing them back to disk. Such sorting is used to facilitate efficient query of k-mer frequencies.} and release the memory occupied by that hash table. Since all the occurrences of the same k-mer will locate in the same partition, the frequency count of a k-mer can be found in only one disk file. This is a very good property, as we do not have to later merge these frequency count disk files. The query of a k-mer's frequency is also very easy and efficient. Given a query k-mer, we can use MSP to calculate its partition location and then perform binary search on the corresponding count disk file to get the k-mer frequency.

\section{Experimental Results}
In this section, we present experimental results that illustrate the efficiency of our MSPKmerCounter on four large real-life short reads data sets: Budgerigar (bird), Red tailed boa constrictor (snake),  Lake Malawi cichlid (fish) and soybean. (1) We first analyze the efficiency of MSPKmerCounter by reporting the memory and time costs, along with the temporary disk space usage; (2) We then investigate the scalability and parallelizability of MSPKmerCounter. We will compare MSPKmerCounter with two state-of-the-art k-mer countering tools: Jellyfish (\citealp{Marcais11}) and BFCounter (\citealp{Melsted11}). All the experiments, if not specifically mentioned, are conducted on a server with 2.00GHz Intel Xeon CPU and 512 GB RAM.

\subsection{Sequence Datasets}
Four very large real-life short reads data sets are used to test MSPKmerCounter. The first one is the sequence data of Budgerigar (bird) obtained from \url{bioshare.bioinformatics.ucdavis.edu/Data/hcbxz0i7kg/Parrot/BGI_illumina_data/}. These short reads were sequenced from the Illumina HiSeq 2000 technology\nop{ for a genome of $\sim$1.23 GB}. The second one is the sequence data of Red tailed boa constrictor (snake) downloaded from \url{bioshare.bioinformatics.ucdavis.edu/Data/hcbxz0i7kg/Snake/short_inserts/}. These short reads were obtained with the Genome Analyzer technology. The third one is the sequence data of Lake Malawi cichlid (fish) downloaded from \url{bioshare.bioinformatics.ucdavis.edu/ Data/hcbxz0i7kg/fish/}. And the last one is the sequence data of soybean downloaded from \url{ftp://public.genomics.org.cn/BGI/soybean_resequencing/fastq/}. Some basic facts about these four data sets are shown in Table \ref{Tab:01}.

\begin{table}[!t]
\processtable{Basic facts about the four sequence data sets used in our experiments\label{Tab:01}}
{\begin{tabular}{lcccc}\toprule
 & bird & snake & fish & soybean \\\midrule
Format &fastq &fastq &fastq &fastq\\
Size (GB) & 106.8 & 181.7 & 137.4 & 40.1\\
Avg Read Length & 150 & 121 & 101 & 75\\
No. of Reads (million) & 323 & 573 & 598 & 227\\\botrule
\end{tabular}}{}
\vspace{-6mm}
\end{table}

\subsection{K-mer Counting Efficiency}
\label{sec:countEff}
We conduct experiments to test the efficiency of our MSPKmerCounter and compare it with two state-of-the-art k-mer counting algorithms: Jellyfish, which is a fast, memory efficient k-mer counting tool based on a multi-threaded, lock-free hash table optimized for counting k-mers up to 31 nucleotides in length; and BFCounter, which is a k-mer counting tool with greatly reduced memory requirements based on bloom filter, a probabilistic data structure. BFCounter is a completely in-memory kmer counting method. Jellyfish can work both as in-memory or out-of-core. It requires user to pre-specify the size of the hash table: if the hash table is large enough to hold all the k-mers, it will be an in-memory method; otherwise, whenever the hash table fills up, the intermediate results will be written to disk and merged later, making it become a disk-based method. In this set of experiments, we will test Jellyfish under these two different settings, denoted as Jellyfish(Memory) and Jellyfish(Disk) respectively. For Jellyfish(Memory), we pre-calculate the number of distinct k-mers in each data set to make sure the hash table is big enough to hold all the k-mers. For Jellyfish(Disk), we set the hash table size to a fixed number so that it will consistently make use of $\sim$11 GB memory. We set the number of threads to 1 for all the three methods, since BFCounter only supports single thread. For MSPKmerCounter, we set the number of wrapped partitions to 1,000 (to reduce memory footprint) and the minimum substring length p to 10.  

Table \ref{Tab:02} presents the memory consumption and running time for the three methods when applied to the snake, fish and soybean data sets\footnote{We reserve the bird data set to test scalability (See Section \ref{sec:scale})} for counting 31-mers\footnote{Jellyfish only supports counting k-mers whose length is smaller than 32. BFCounter and MSPKmerCounter do not have such a constraint.}.

\begin{table}[!th]
\processtable{Comparison of memory consumption and running time for counting 31-mers on the snake, fish and soybean data sets. \label{Tab:02}}
{\begin{tabular}{lccccccc}\toprule
Algorithm & \multicolumn{3}{c}{Memory (GB)} & & \multicolumn{3}{c}{Run Time (minutes)} \\
\cline{2-4} \cline{6-8} \\
          & snake & fish & soybean & & snake & fish & soybean \\\midrule
Jellyfish(Memory) & 110 & 114 & 43 & & 455.5 & 374.5 & 93.6 \\
Jellyfish(Disk) & 11 & 11 & 11 & & 775.2 & 503.7 & 117.7 \\
BFCounter & 38 & 29 & 13 & & 1899.8 & 1299 & 342.2\\
MSPKmerCounter & 9.6 & 9.9 & 6.3 & & 492.7 & 399.2 & 99\\\botrule
\end{tabular}}{}
\vspace{-3mm}
\end{table}

Table \ref{Tab:03} shows the temporary disk space usage for the three methods when applied to the snake, fish and soybean data sets for counting 31-mers.

\begin{table}[!h]
\processtable{Comparison of temporary disk space usage for counting 31-mers on the snake, fish and soybean data sets. \label{Tab:03}}
{\begin{tabular}{lccc}\toprule
Algorithm & \multicolumn{3}{c}{Temp Disk Space Usage (GB)} \\
\cline{2-4}  \\
          & snake & fish & soybean \\\midrule
Jellyfish(Memory) & 0 & 0 & 0 \\
Jellyfish(Disk) & 332 & 197 & 44 \\
BFCounter & 0 & 0 & 0 \\
MSPKmerCounter & 217 & 168 & 43 \\\botrule
\end{tabular}}{}
\vspace{-3mm}
\end{table}

As can be seen from Table \ref{Tab:02}, when applied to a large sequence data set with deep coverage, our MSPKmerCounter soon demonstrates its advantages. It uses much less memory than both Jellyfish(Memory) and BFCounter. Its running time is close to that of Jellyfish(Memory) and significantly shorter than that of BFCounter. Jellyfish(Disk) was able to finish the counting task using a small amount of memory, by writing intermediate results to disk and later merging them. But unfortunately, its merging process is relatively inefficient since the k-mer sets in those intermediate results are not completely disjoint. Therefore it is much slower than MSPKmerCounter. MSPKmerCounter requires no additional merging steps after partial counting results are generated from individual partitions. Also, as can be seen from Table \ref{Tab:03}, MSPKmerCounter uses less amount of temporary disk space than Jellyfish(Disk). Note that Jellyfish(Memory) and BFCounter do not need to use any temporary disk space since they are completely memory-based.  Actually the memory consumption and temporary disk space usage of our MSP-based counting method can be fully controlled by varying the number of wrapped partitions and the minimum substring length $p$. For more discussions (both theoretical and experimental) about the sensitivity of MSP to these parameters, please refer to \cite{Li13}.

\subsection{Scalability}
\label{sec:scale}

We then conduct experiments to test the scalability of Jellyfish, BFCounter and our MSPKmerCounter. Specifically, we count the k-mers in the Budgerigar data set for various levels of coverage, using these three counting methods. In order to get different levels of coverage, we randomly sampled the short reads data set to obtain a desired amount of sequences. As same as the previous experiments, here we also test Jellyfish under two different settings.

The memory consumption, running time and temporary disk space usage for counting 31-mers in the Budgerigar(bird) data set under various levels of coverage are shown in Figures \ref{fig:mem}, \ref{fig:time} and \ref{fig:disk}, respectively.

\begin{figure*}[t!]
\centering
\subfigure[\small\textit{Memory Consumption}]{\label{fig:mem}\includegraphics[scale=0.25]{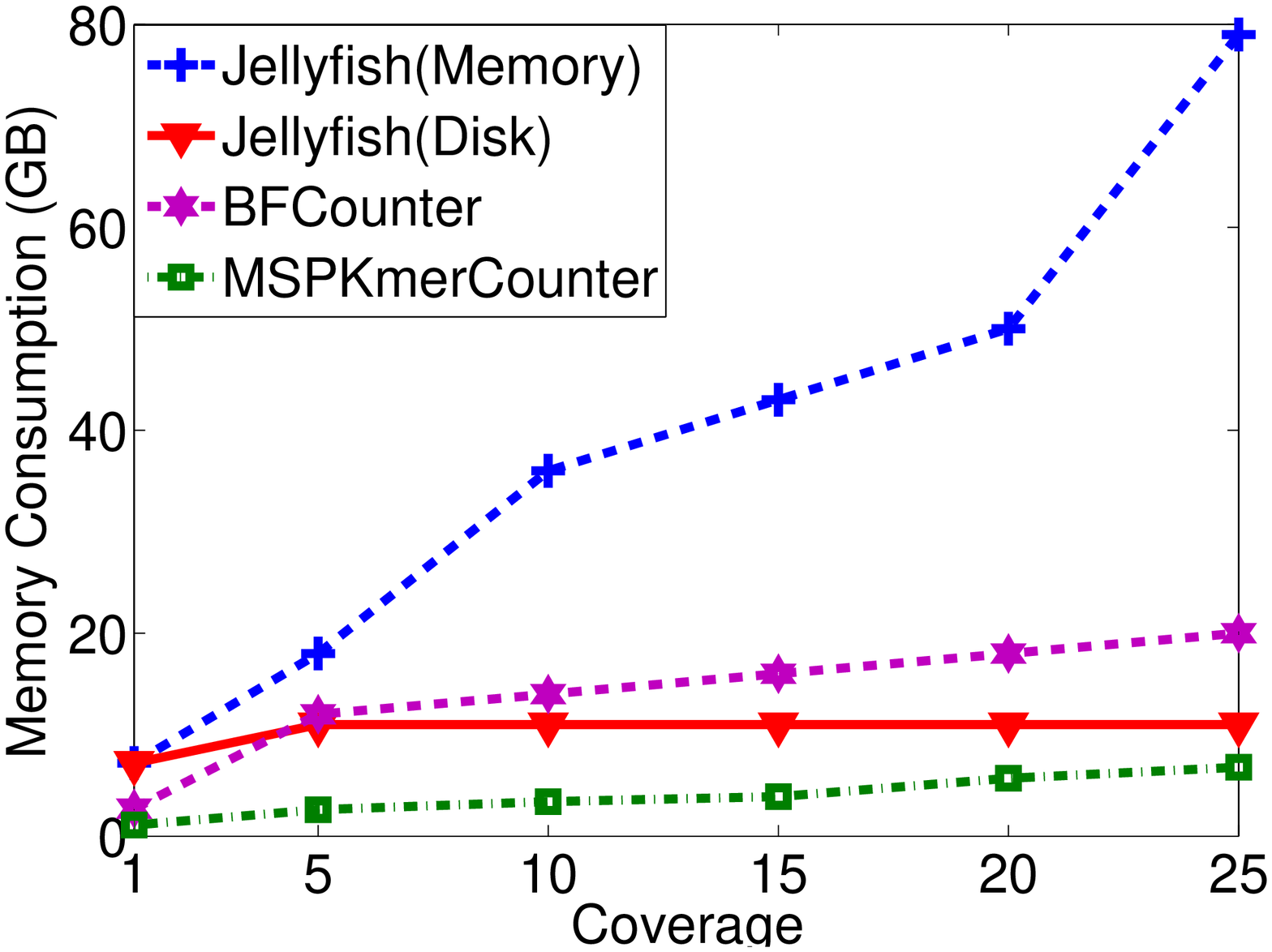}}
\subfigure[\small\textit{Running Time}]{\label{fig:time}\includegraphics[scale=0.25]{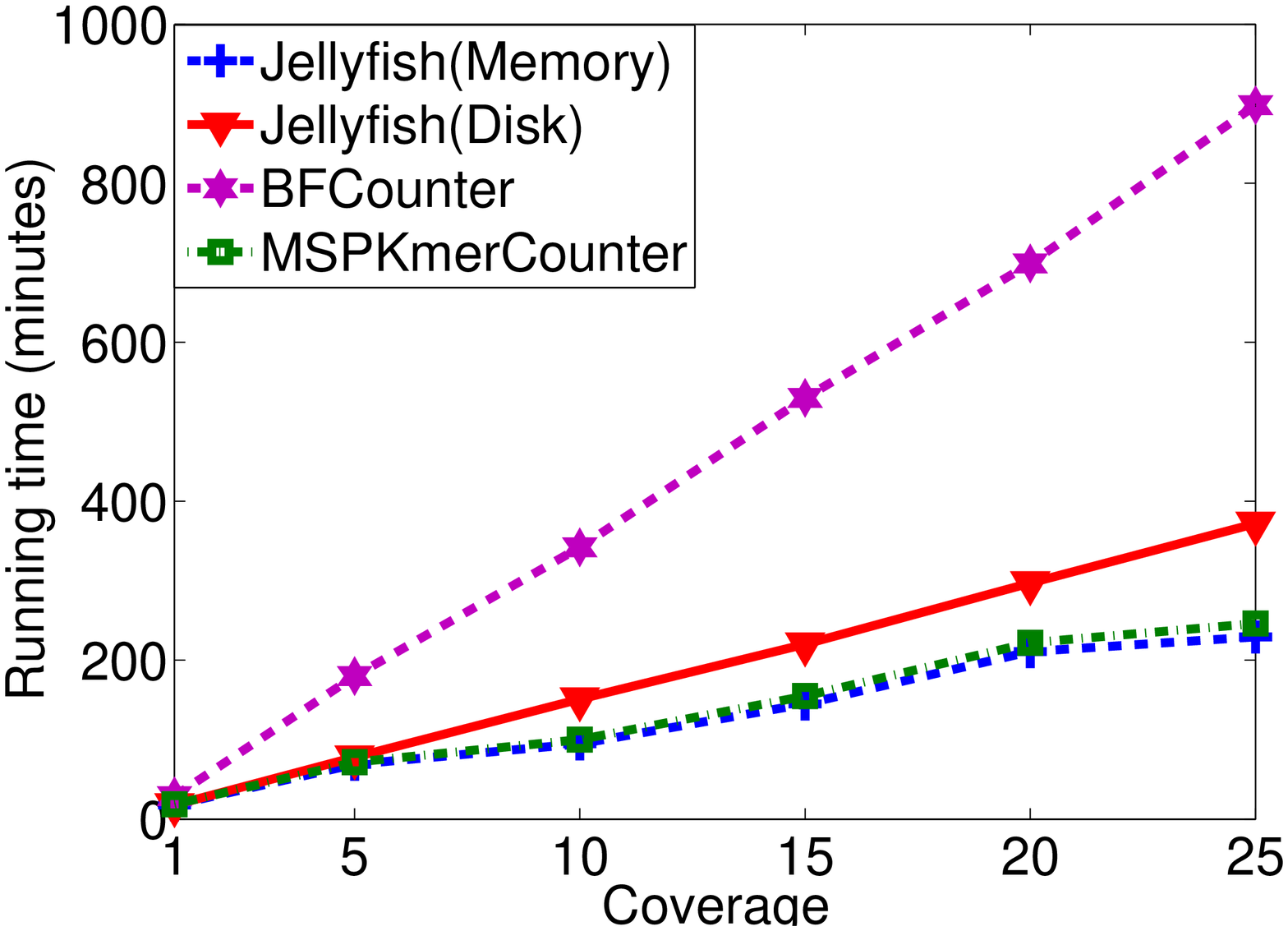}}
\subfigure[\small\textit{Temp Disk Space Usage}]{\label{fig:disk}\includegraphics[scale=0.25]{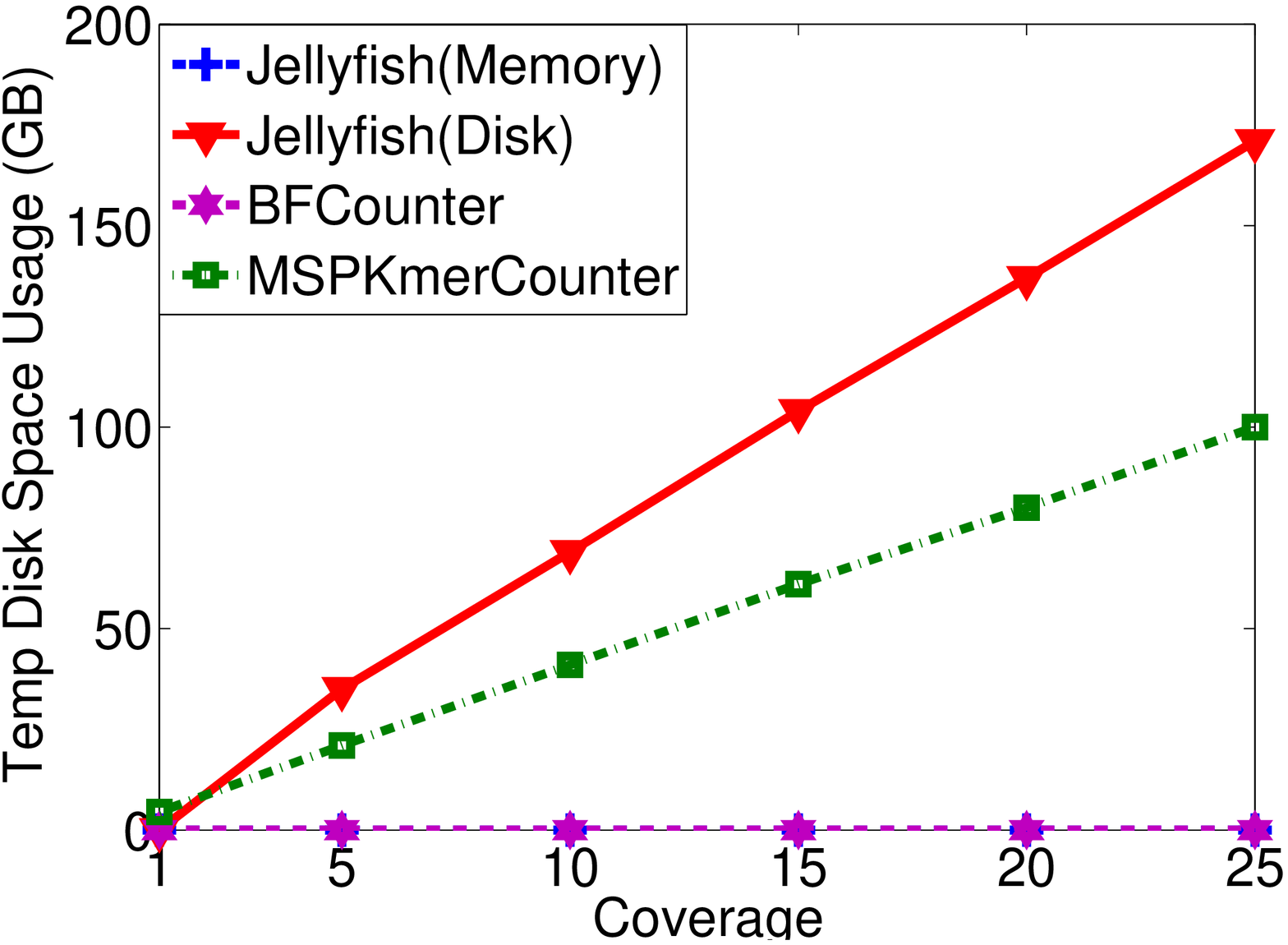}}
\caption{Memory consumption, running time and temporary disk space usage of Jellyfish, BFCounter and MSPKmerCounter for counting 31-mers in the Budgerigar data set under various levels of coverage}
\label{fig:plength}
\end{figure*}

\nop{
\begin{figure}[!tpbh]
\centerline{\includegraphics[width=0.45\textwidth]{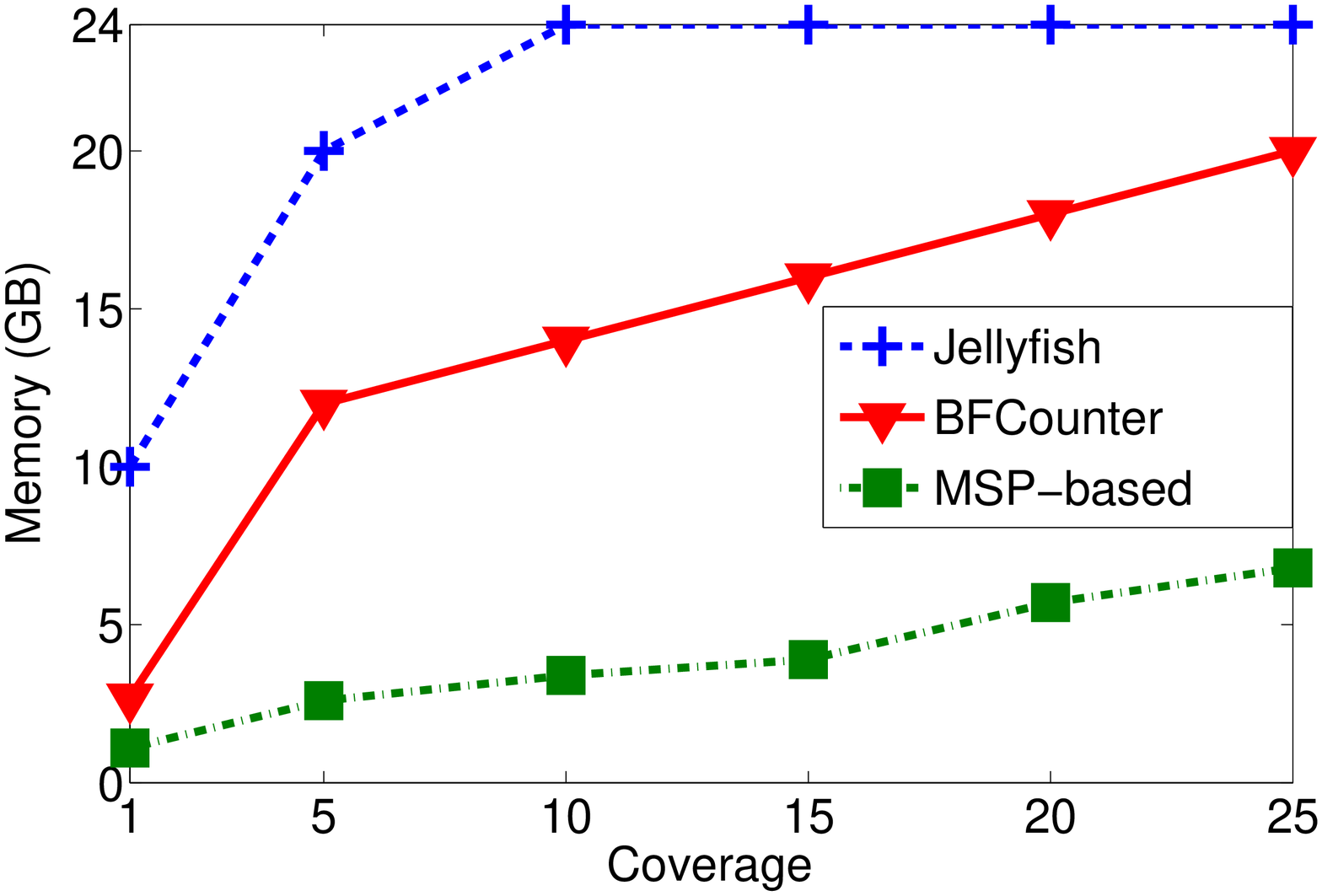}}
\vspace{-5mm}
\caption{Memory consumption of Jellyfish, BFCounter and MSPKmerCounter for counting 31-mers in the Budgerigar data set under various levels of coverage}\label{fig:mem}
\end{figure}

\begin{figure}[!tpbh]
\centerline{\includegraphics[width=0.45\textwidth]{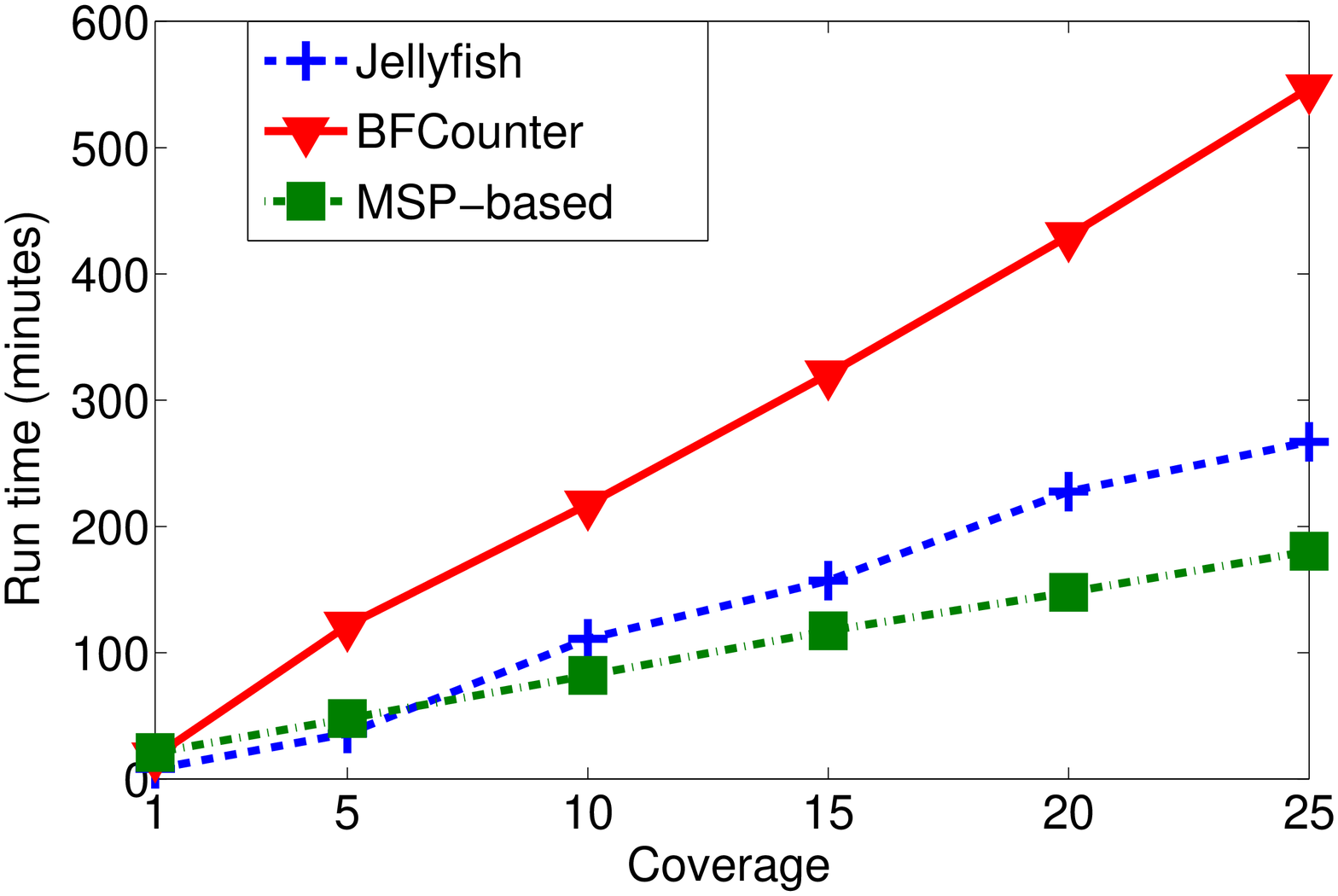}}
\vspace{-5mm}
\caption{Running time of Jellyfish, BFCounter and MSPKmerCounter for counting 31-mers in the Budgerigar data set under various levels of coverage}\label{fig:time}
\end{figure}

\begin{figure}[!tpbh]
\centerline{\includegraphics[width=0.45\textwidth]{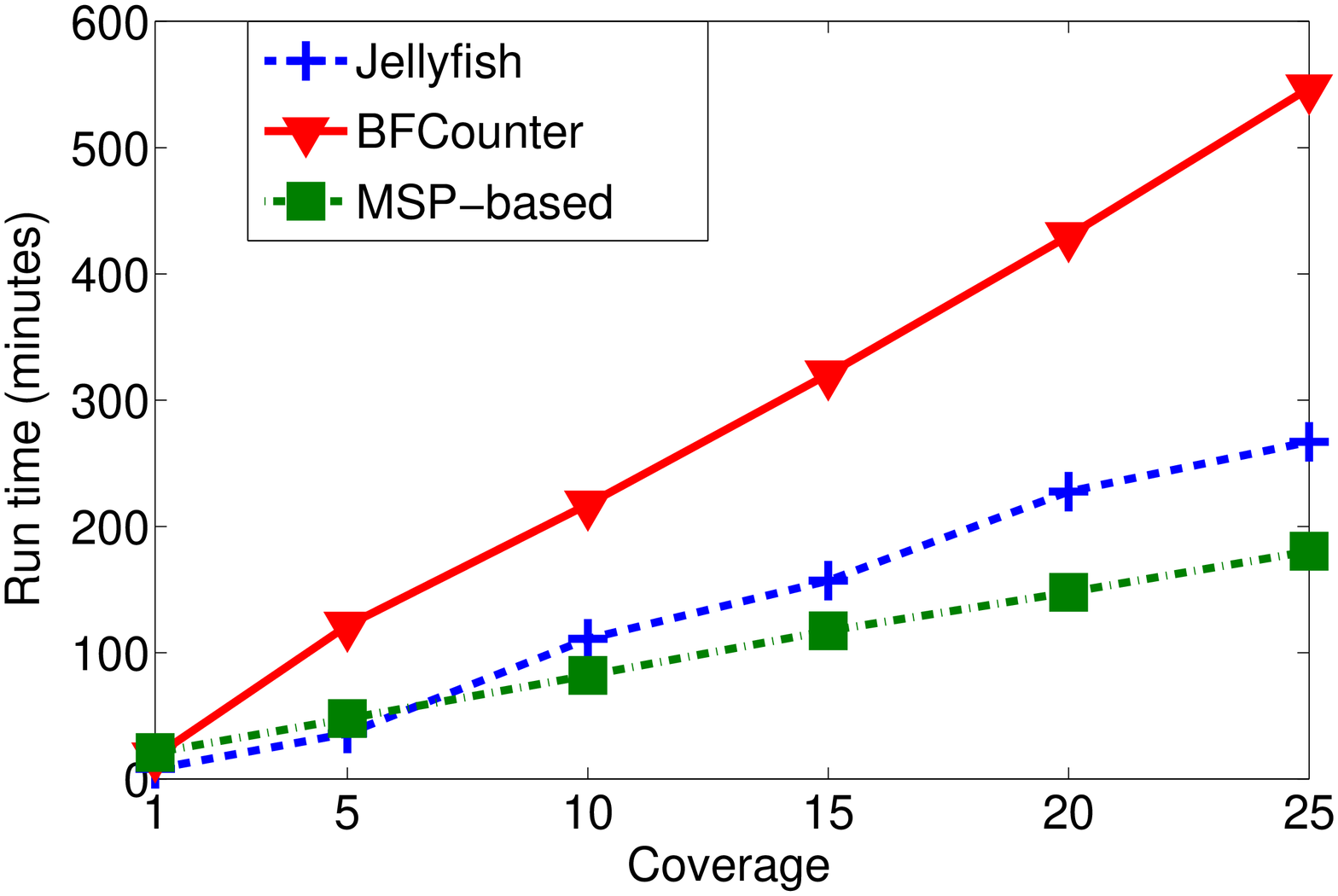}}
\vspace{-5mm}
\caption{Temporary disk space usage of Jellyfish, BFCounter and MSPKmerCounter for counting 31-mers in the Budgerigar data set under various levels of coverage}\label{fig:disk}
\end{figure}
}

Figures \ref{fig:mem} shows that with the increase of coverage, the memory consumption of Jellyfish(Memory) increases significantly. In comparison, the memory utilizations of BFCounter and MSPKmerCounter only increase slightly. MSPKmerCounter outperforms both Jellyfish and BFCounter in terms of memory footprint. Note that we configure Jellyfish(Disk) to use at most 11 GB memory, so its memory consumption does not change since coverage 5.  Figure \ref{fig:time} shows that with the increase of coverage, the running time of all counting methods increases. However, the increasing speed of BFCounter is much higher than that of Jellyfish and MSPKmerCounter. As the coverage increases, the running time gap between MSPKmerCounter and Jellyfish(Disk) becomes larger and larger, indicating MSPKmerCounter's better scalability. Even when compared with the purely memory-based Jellyfish(Memory), MSPKmerCounter is only slightly slower at all coverages. Figure \ref{fig:disk} shows that the temporary disk space usages of both Jellyfish(Disk) and MSPKmerCounter increase as the coverage increases. But the increasing speed of MSPKmerCounter is much slower than that of Jellyfish(Disk), indicating MSPKmerCounter's better scalability in disk space utilization. Jellyfish(Memory) and BFCounter need no extra disk space since they are completely memory-based. To summary, when the coverage is low (e.g. less than 5), the performance differences among Jellyfish, BFCounter and MSPKmerCounter are not very big, though MSPKmerCounter is still much faster than BFCounter and uses less memory than both Jellyfish and BFCounter. As the coverage increases, MSPKmerCounter quickly dominates the scene. In a high coverage situation, the main memory is not big enough for Jellyfish to finish all the computation in memory and therefore it has to write intermediate results to disk and later merge them. This gives MSPKmerCounter a chance to outperform Jellyfish in terms of both memory and time. BFCounter has the advantage that its memory usage does not increase a lot as the coverage increases. However, compared with MSPKmerCounter, it still requires more memory and much longer running time to finish the task. Moreover, the memory consumption, running time and disk space usage of MSPKmerCounter are fully controllable by varying several parameters (\citealp{Li13}).

\subsection{Parallelizability}
Our MSP-based k-mer counting method can easily be parallelized to support multi-threads or be distributed to multiple machines to enable parallel processing. There are three distinct phases in the Minimum Substring Partitioning process. First, it reads the short read sequences. Second, it calculates the minimum substring of each k-mer and merges possible adjacent k-mers into ``super k-mers".  Last, it writes the ``super k-mers" back to disk files. Phase 1 and phase 3 are I/O operations, so the speedup can be obtained by using multi-threads to process phase 2. After partitioning, different partitions are completely disjoint. Therefore it is helpful to use different threads to process different partitions simultaneously. 

\begin{figure}[!h]
\centerline{\includegraphics[width=0.4\textwidth]{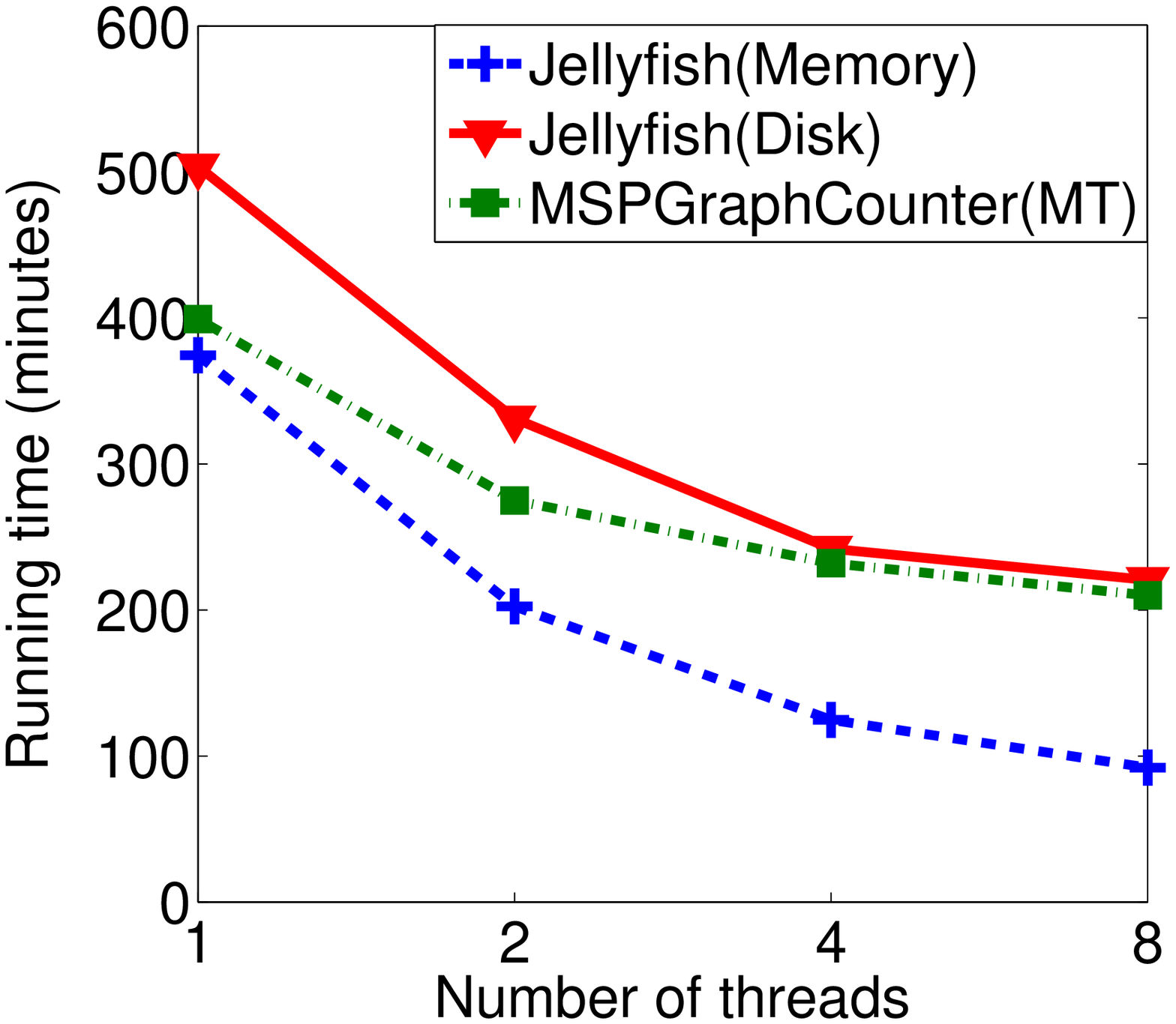}}
\caption{Running time versus \#threads for Jellyfish and MSPKmerCounter}\label{fig:multithread}
\vspace{-5mm}
\end{figure}

We implemented a preliminary multi-thread version of our MSPKmerCounter, denoted as MSPKmerCounter(MT). Since BFCounter is not able to support multiple threads, here we only conduct experiments to compare MSPKmerCounter(MT) with Jellyfish, which is highly optimized to support efficient multi-thread processing. Figure \ref{fig:multithread} shows the running time comparison of MSPKmerCounter(MT) and Jellyfish with the increasing number of threads.  Here k-mers are counted on the Lake Malawi cichlid (fish) data set with $k=31$. Again we test Jellyfish under two different settings (the settings are as same as those in Section \ref{sec:countEff}). From Figure \ref{fig:multithread} we can see that: (1) Jellyfish(Memory) has an almost linear speedup up to 4 threads, indicating the best parallelizability. This is reasonable since it puts everything in memory and therefore involves almost no I/O costs. However, as mentioned before, its huge memory footprint will greatly limit its usage on commodity computers. (2) Both Jellyfish(Disk) and MSPKmerCounter(MT) exhibit good parallelizability for up to 2 threads and then levels off. This is because these two disk based methods involve a lot of I/O operations. At 2 threads the CPU calculation is already fast enough and the I/O bandwidth has become the main bottleneck. MSPKmerCounter(MT) is still faster than Jellyfish(Disk).

From (1) and (2) we can conclude that Jellyfish is more suitable for powerful computers (e.g. computers with large RAM and many cores), while MSPKmerCounter is the better choice for commodity computers (e.g. computers with small RAM and few cores).

\vspace{-2mm}

\section{Future Work}
There are some future avenues to pursue to further improve our work. First, we can adopt the techniques (e.g. variable length encoding) introduced in Jellyfish (\citealp{Marcais11}) to make space-efficient encoding of keys and reduce the memory usage of each hash entry to further reduce the memory consumption. Second, we can think about extending the use of MSP from counting k-mers to the whole sequence assembly process. Since the k-mers in different MSP partitions are completely disjoint and the majority of adjacent k-mers in original short reads are retained in the same partition, it is possible to perform local assembly (including some error correction steps like tip removal and bubble merging) for each partition and later ``glue" these local assembly results to obtain the global assembly results. By doing so, the whole assembly process can be done with a very small amount of memory. And the assembly can speed up a lot with the gains of parallel assembly of multiple partitions.

\section{Conclusion}
In this paper, we aimed at the computational bottlenecks in k-mer counting, which is an important step in many genome sequence assembly tasks. We developed a disk-based approach based on a novel technique, Minimum Substring Partitioning (MSP), to solve the memory overwhelming problem.  MSP breaks the short reads into multiple disjoint partitions so that each partition only requires a very small amount of memory to process.  By leveraging the overlaps among the k-mers derived from the same read, MSP is able to achieve astonishing compression ratio so that the I/O cost can be greatly reduced, making the method be very efficient in terms of time and space. Our MSP-based k-mer counting method were evaluated on real DNA short read sequences. Experimental results show that it can not only successfully finish the counting task on very large data sets using a reasonable amount of memory, but also achieve better overall performance than the existing k-mer counting methods.

\section*{Acknowledgement}
 The authors would like to thank the Assemblathon website (\url{http://assemblathon.org}) for providing the Budgerigar (bird), the Red tailed boa constrictor (snake) and the Lake Malawi cichlid (fish) data, and Beijing Genomics Institute for providing the soybean data.

\paragraph{Funding\textcolon} This research was sponsored in part by the U.S. National Science Foundation under grant IIS-0847925 and IIS-0954125. The views and conclusions contained herein are those of the authors and should not be interpreted as representing the official policies, either expressed or implied, of the U.S. Government. The U.S. Government is authorized to reproduce and distribute reprints for Government purposes notwithstanding any copyright notice herein.

%\bibliographystyle{natbib}
%\bibliographystyle{achemnat}
%\bibliographystyle{plainnat}
%\bibliographystyle{abbrv}
%\bibliographystyle{bioinformatics}
%
%\bibliographystyle{plain}
%
%\bibliography{Document}

\begin{thebibliography}{}
\bibitem[Mardis {\it et~al}., 2008]{Mardis08} Mardis, E.R. (2008) Next-generation DNA sequencing methods, {\it Annu. Rev. Genomics Hum. Genet.}, {\bf 9}, 387-402.

\bibitem[Myers {\it et~al}., 2000]{Myers00} Myers, E.W. {\it et~al}. (2000) A whole-genome assembly of Drosophila, {\it Science}, {\bf 287}, 2196-2204.

\bibitem[Batzoglou {\it et~al}., 2002]{Batzoglou02} Batzoglou, S. {\it et~al}. (2002) ARACHNE: a whole-genome shotgun assembler, {\it Genome research}, {\bf 12}, 177-189.

\bibitem[Havlak {\it et~al}., 2004]{Havlak04} Havlak, P. {\it et~al}. (2004) The Atlas genome assembly system, {\it Genome research}, {\bf 14}, 721-732.

\bibitem[Mullikin {\it et~al}., 2003]{Mullikin03} Mullikin, J.C. and Ning, Z. (2003) The phusion assembler, {\it Genome research}, {\bf 13}, 81-90.

\bibitem[Platt {\it et~al}., 2010]{Platt10} Platt, D. and Evers, DJ (2010) Forge: A Parallel Genome Assembler Combining Sanger and Next Generation Sequence Data, \textit{http://combiol.org/forge/}.

\bibitem[Pevzner {\it et~al}., 2001]{Pevzner01} Pevzner, P.A. {\it et~al}. (2001) An Eulerian path approach to DNA fragment assembly, {\it Proceedings of the National Academy of Sciences}, 9748-9753.

\bibitem[Zerbino {\it et~al}., 2008]{Zerbino08} Zerbino, D.R. and Birney, E. (2008) Velvet: algorithms for de novo short read assembly using de Bruijn graphs, {\it Genome research}, {\bf 18}, 821-829.

\bibitem[Simpson {\it et~al}., 2009]{Simpson09} Simpson, J.T. {\it et~al}. (2009) ABySS: a parallel assembler for short read sequence data, {\it Genome research}, {\bf 19}, 1117-1123.

\bibitem[Butler {\it et~al}., 2008]{Butler08} Butler, J. {\it et~al}. (2008) ALLPATHS: de novo assembly of whole-genome shotgun microreads, {\it Genome research}, {\bf 18}, 810-820.

\bibitem[Li {\it et~al}., 2010a]{Li10a} Li, R. {\it et~al}. (2010) De novo assembly of human genomes with massively parallel short read sequencing, {\it Genome research}, {\bf 20}, 265-272.

\bibitem[Miller {\it et~al}., 2010]{Miller10} Miller, J.R. {\it et~al}. (2010) Assembly algorithms for next-generation sequencing data, {\it Genomics}, {\bf 95}, 315-327.

\bibitem[Li {\it et~al}., 2010b]{Li10b} Li, R. {\it et~al}. (2010) The sequence and de novo assembly of the giant panda genome, {\it Nature}, {\bf 463(7279)}, 311-317.

\bibitem[Marcais {\it et~al}., 2011]{Marcais11} Marcais, G. and Kingsford, C. (2011) A fast, lock-free approach for efficient parallel counting of occurrences of k-mers, {\it Bioinformatics}, {\bf 27}, 764-770.

\bibitem[Melsted {\it et~al}., 2011]{Melsted11} Melsted, P. and Pritchard, J.K. (2011) Efficient counting of k-mers in DNA sequences using a bloom filter, {\it BMC Bioinformatics}, {\bf 12}, 333-339.

\bibitem[Salzberg {\it et~al}., 2012]{Salzberg12} Salzberg, S.L. {\it et~al}. (2012) GAGE: A critical evaluation of genome assemblies and assembly algorithms, {\it Genome research}.

\bibitem[Earl {\it et~al}., 2011]{Earl11} Earl, D. {\it et~al}. (2011) Assemblathon 1: A competitive assessment of de novo short read assembly methods, {\it Genome research}, {\bf 21}, 2224-2241.

\bibitem[Iqbal {\it et~al}., 2012]{Iqbal12} Iqbal, Z. {\it et~al}. (2012) De novo assembly and genotyping of variants using colored de Bruijn graphs, {\it Nature genetics}.

\bibitem[Li {\it et~al}., 2013]{Li13} Li, Y. {\it et~al}. (2013) Memory efficient minimum substring partitioning, {\it Proceedings of the Very Large Databases Endowment}, 169-180.


\end{thebibliography}

\end{document}